%% file: main.tex
\documentclass[sigconf]{acmart}

\setcopyright{none}              
\renewcommand\footnotetextcopyrightpermission[1]{} 

\usepackage{algorithm}
\usepackage{algorithmic} 
\usepackage{graphicx}
\usepackage{textcomp}
\usepackage{xcolor}
\usepackage{tabularx}
\usepackage{multirow}
\usepackage{float}
\usepackage{multicol}
\usepackage{booktabs}
\usepackage{makecell}
\usepackage{balance}
\usepackage{comment}
\usepackage{enumitem}

\begin{document}



\title{Spec-Driven Hardware Evolution via Executable Contract Refinement and Proof-Guided RTL Update}



\author{Shibo Zhao}
\affiliation{
    \institution{Southeast University}
    \city{Nanjing}
    \country{China}
}

\author{Yang Zhang}
\affiliation{
    \institution{Southeast University}
    \city{Nanjing}
    \country{China}
}

\author{Mengxia Tao}
\affiliation{
    \institution{National Center of Technology Innovation for EDA}
    \city{Nanjign}
    \country{China}
}

\author{Baoqi Zhang}

\affiliation{
    \institution{National Center of Technology Innovation for EDA}
    \city{Nanjign}
    \country{China}
}

\author{Kezhi Li}
\affiliation{
    \institution{The Chinese University of Hong Kong}
    \city{Hong Kong}
    \country{China}
}

\author{Qiang Xu}
\affiliation{
    \institution{The Chinese University of Hong Kong}
    \city{Hong Kong}
    \country{China}
}

\author{Binwu Zhu}
\affiliation{
    \institution{Southeast University}
    \city{Nanjing}
    \country{China}
}

\author{Hao Yan}
\affiliation{
    \institution{Southeast University}
    \city{Nanjing}
    \country{China}
}

\author{Min Li}
\affiliation{
    \institution{Southeast University}
    \city{Nanjing}
    \country{China}
}

\begin{abstract}
\input{tex/abstract}

\end{abstract}

\maketitle
\pagestyle{plain}                
\input{tex/introduction}

\input{tex/background}

\input{tex/problem}

\input{tex/methodology}

\input{tex/experiment}
\input{tex/conclusion}

{
\balance
\bibliographystyle{ACM-Reference-Format}
\bibliography{./references}
}

\end{document}

%% file: tex/abstract.tex
Hardware development is inherently evolutionary: major revisions typically begin by changing intended behavior and then updating a previously validated implementation, rather than regenerating RTL from scratch. Yet most recent LLM-based hardware research still frames the task primarily as prompt-to-RTL generation, offering limited support for semantic version evolution of trusted legacy designs.
We present \emph{spec-driven hardware evolution}, a contract-centered formulation for RTL version iteration. Instead of treating a new feature request as a direct prompt for RTL generation, we refine it into a reviewed \emph{executable contract} for the next version. This contract specifies \emph{what} must hold at the externally visible transactional level through a behavior-level reference together with explicit observation and checking semantics, while leaving \emph{how} the change is realized in RTL to the evolution process. Based on this formulation, we organize hardware evolution into four stages: \emph{Specify}, \emph{Plan}, \emph{Implement}, and \emph{Validate}. After contract approval, the remaining stages proceed automatically: \emph{Plan} derives cross-version semantic deltas and localizes affected RTL regions, aided by \emph{mutation-based semantic probing}; \emph{Implement} and \emph{Validate} then perform legacy-aware RTL update under proof-guided checking and iterative repair.
We evaluate the framework on a controlled version-evolution case study of a representative TPU datapath block under data-format changes. The results support the feasibility of contract-driven hardware evolution and demonstrate that the proposed backend workflow can effectively drive validated legacy RTL toward next-version functional convergence under a reviewed executable contract. An anonymous artifact for reproducibility is available at \url{https://anonymous.4open.science/r/SDHE-3A6C}.

%% file: tex/introduction.tex
\section{Introduction}

Recent progress in LLM-based coding agents and hardware-oriented language models substantially advances RTL generation, code completion, and hardware-design assistance~\cite{liu2024rtlcoder,thakur2024verigen,gao2024autovcoder,cui2024origen,pei2024betterv,zhao2025codev,zhu2025codevr1}. Public benchmarks further improve the evaluation of hardware-code generation and reasoning systems~\cite{liu2023verilogeval,liu2024openllm,jin2025realbench}, while emerging agent frameworks begin to explore richer specifications, repository-scale contexts, and tool-assisted workflows for hardware design~\cite{li2025specllm,ho2025verilogcoder,zhao2025mage,wei2025vflow,allam2025asic,tasnia2025veriopt}. Despite this progress, most existing formulations still center on prompt-to-RTL generation, standalone RTL completion, or direct RTL editing from informal task descriptions. These abstractions are valuable for studying code generation, but they only partially capture practical hardware maintenance, where engineers more often evolve a validated legacy design across versions than regenerate a new implementation from scratch.

This distinction matters because version evolution is a central mode of real hardware development. Across releases, engineers revise functionality, interface behavior, corner-case handling, and selected timing expectations while attempting to preserve as much validated logic as possible. For nontrivial updates, the first artifact that must be stabilized is often not the RTL itself, but the intended next-version behavior: what should change, under what interface conditions, and with what observation rules. This motivates a hardware analogue of \emph{spec-driven development}~\cite{piskala2026spec}, but under a stricter trust boundary: in hardware, a natural-language request alone is rarely sufficient, because correctness must ultimately be established through simulation and formal verification~\cite{blum2002reflections}. The semantic target for evolution must therefore be \emph{reviewable}, \emph{executable}, and \emph{verification-consumable}~\cite{binkert2011gem5,clarke2003behavioral}.

In this paper, we capture that target as an \emph{executable contract}. An executable contract consists of a behavior-level reference together with explicit observation and checking semantics for externally visible transactions. It defines \emph{what} the next version must satisfy, without prescribing \emph{how} the change should be realized in RTL. Under this formulation, the reference serves as a semantic target for evolution rather than an implementation recipe. The setting therefore differs fundamentally from both informal prompt-based RTL editing and behavioral-design flows such as high-level synthesis (HLS), which treat high-level code as a source for implementation generation~\cite{lahti2018we}.

Based on this view, we formulate \emph{spec-driven hardware evolution} as a four-stage workflow: \emph{Specify}, \emph{Plan}, \emph{Implement}, and \emph{Validate}. In \emph{Specify}, a new feature request is refined into a candidate executable contract for version-$N{+}1$ and explicitly reviewed before downstream evolution proceeds. Once approved, \emph{Plan} derives semantic deltas and localizes the RTL regions most relevant to the intended change, \emph{Implement} incrementally updates the validated legacy RTL under this guidance, and \emph{Validate} checks the candidate RTL against the executable contract using proof-oriented checking and/or co-simulation, with failures fed back into iterative repair. Together, these stages turn hardware evolution into a contract-driven, proof-guided process over a legacy-anchored design space.

Realizing this workflow raises several technical challenges beyond the workflow formulation itself. First, the next-version contract must be made precise enough for downstream checking: the intended behavior, interface conditions, observation rules, and timing semantics must be rendered into a reviewed executable artifact rather than left implicit in natural language. Second, once the contract changes, the system must determine how the revised behavior relates to the validated legacy RTL: which behaviors should be preserved, where the semantic delta lies, and which RTL regions are most likely to require modification. Third, localized RTL edits must be driven to functional closure under a verification signal strong enough to distinguish true semantic progress from superficial code changes. To address these difficulties, we instantiate the framework with three corresponding mechanisms: executable-contract refinement in the front end, \emph{mutation-based semantic probing} for semantic alignment and change localization in \emph{Plan}, and a \emph{proof-guided RTL update} loop that uses formal-checking outcomes as structured feedback for localized repair and contract satisfaction.

Our current focus is \emph{functional evolution} rather than PPA optimization. Accordingly, the RTL produced by the proposed flow is intended to provide a functionally converged starting point for downstream PPA-oriented refinement~\cite{pan2024physically}. This scope aligns well with practical maintenance scenarios, whereas existing academic benchmarks~\cite{liu2023verilogeval,liu2024openllm,jin2025realbench} for LLM-based Verilog generation or RTL editing mainly target prompt-to-code generation, standalone RTL completion, or repository-level editing rather than contract-driven version evolution over trusted legacy artifacts. We therefore evaluate the proposed flow on controlled version evolution of a representative TPU datapath block~\cite{jouppi2017datacenter} under data-format changes~\cite{valero2023mixed}.

The contributions of this paper are as follows:
\begin{itemize}
\item We formulate \emph{spec-driven hardware evolution} as a contract-centered setting for semantic RTL version evolution, organized into four stages---\emph{Specify}, \emph{Plan}, \emph{Implement}, and \emph{Validate}---that combine reviewed next-version intent, legacy-aware change planning, incremental RTL update, and proof-guided validation.

\item We develop a hardware-specific front-end flow for executable contract refinement and semantic change analysis, which turns feature requests into reviewed next-version contracts and derives cross-version behavioral deltas against validated legacy designs.

\item We introduce \emph{mutation-based semantic probing} and a \emph{proof-guided RTL update} loop for automated hardware evolution: the former establishes grounded contract-to-RTL correspondences and localizes change-relevant regions in the legacy design, while the latter performs localized legacy-aware implementation updates under formal-feedback-driven repair until the revised contract is satisfied.
\end{itemize}

%% file: tex/background.tex
\section{Related Work}
\label{related_work}

\subsection{Software Engineering Agents and Executable Specifications}

Recent advances in software engineering agents shift the focus from isolated code generation to long-horizon, repository-level workflows. Agents such as CodeAgent~\cite{zhang2024codeagent} and SWE-agent~\cite{yang2024swe}, together with benchmarks such as SWE-bench~\cite{jimenez2023swe} and SWE-EVO~\cite{thai2025swe}, emphasize iterative problem solving over existing codebases. Systems such as InterCode~\cite{yang2023intercode}, ChatRepair~\cite{xia2024automated}, and RepairAgent~\cite{bouzenia2025repairagent} further demonstrate the value of closed-loop execution with tool interaction and environment feedback.

At the same time, recent work elevates specifications from informal descriptions to executable artifacts. Frameworks such as SysSpec~\cite{liu2026sharpen} and SpecGen~\cite{ma2025specgen} refine specifications into semantically clearer and more checkable forms. However, in software settings these artifacts typically improve prompting, testing, or repair, rather than serve as binding contracts that define the target behavior of a version-evolution process.

These trends are conceptually relevant to our work, but hardware imposes a stricter correctness regime. In RTL evolution, correctness cannot rely only on empirical execution outcomes, because cycle-level behavior, interface timing, and sequential dependencies often require formal reasoning. This motivates our use of reviewed executable contracts not as auxiliary prompts, but as the semantic target that governs hardware evolution.

\subsection{LLM-Based RTL Generation and RTL Evolution}

A growing body of work studies LLMs for RTL generation, code completion, and hardware-design assistance. Benchmarks such as VerilogEval~\cite{liu2023verilogeval} and RTLLM~\cite{lu2024rtllm} establish early public evaluation settings for prompt-to-RTL generation, while more recent efforts such as RealBench~\cite{jin2025realbench} move toward more realistic IP-level tasks with structured specifications and stronger verification.

Most of this literature adopts a de novo generation viewpoint: given a natural-language description, the model produces a functionally correct RTL module from scratch. Fine-tuned models such as RTLCoder~\cite{liu2024rtlcoder} and OriGen~\cite{cui2024origen}, as well as multi-agent systems such as MAGE~\cite{zhao2025mage} and VerilogCoder~\cite{ho2025verilogcoder}, steadily improve generation quality within this setting. However, their starting point is still typically an empty or incomplete design, and the specification remains primarily an informal task description. As a result, these methods do not directly address the setting in which a revised executable contract drives the semantic evolution of a trusted legacy RTL implementation.

A separate line of work considers the evolution of existing RTL designs. EvoVerilog~\cite{guo2025evoverilog} and VeriOpt~\cite{tasnia2025veriopt} focus on PPA-oriented improvement, while CktEvo~\cite{shi2026cktevo} studies repository-level RTL refactoring under functional-equivalence constraints. These works move closer to realistic maintenance scenarios, but their objective remains different from ours: they optimize or refactor existing RTL, rather than evolve it under a reviewed next-version contract that explicitly redefines intended behavior. Our setting is therefore neither prompt-to-code generation nor function-preserving structural refactoring, but contract-driven \emph{functional} version evolution over legacy hardware artifacts.

\subsection{Verification Feedback in LLM-Based Hardware Design}

Recent LLM-based methods incorporate verification feedback with different levels of strength. AutoChip~\cite{blocklove2025automatically} uses simulation-based feedback, VeriReason~\cite{wang2025verireason} exploits testbench outcomes through reinforcement learning, and AssertLLM~\cite{yan2025assertllm} generates verification assertions from natural-language specifications. More recent systems further integrate formal reasoning into the loop: FormalRTL~\cite{li2026formalrtl} uses equivalence checking against reference models with counterexample-guided debugging, HDLFORGE~\cite{abdollahi2026hdlforge} converts BMC traces into reusable micro-tests, and Wit-HW~\cite{ma2025wit} uses mutation-based witness generation for fault localization.

Despite these advances, verification in prior work primarily serves to validate or repair candidate implementations in de novo synthesis or code-generation settings. In contrast, our framework uses formal feedback as an organizing mechanism for \emph{version evolution}: it helps characterize how legacy functionality is realized, localize the impact of a semantic delta, and drive contract-guided RTL update toward next-version convergence. In this sense, verification is not merely a post hoc filter or repair trigger, but a central component of legacy-aware hardware evolution.

%% file: tex/problem.tex
\begin{figure}[t]
    \centering
    \includegraphics[width=\linewidth]{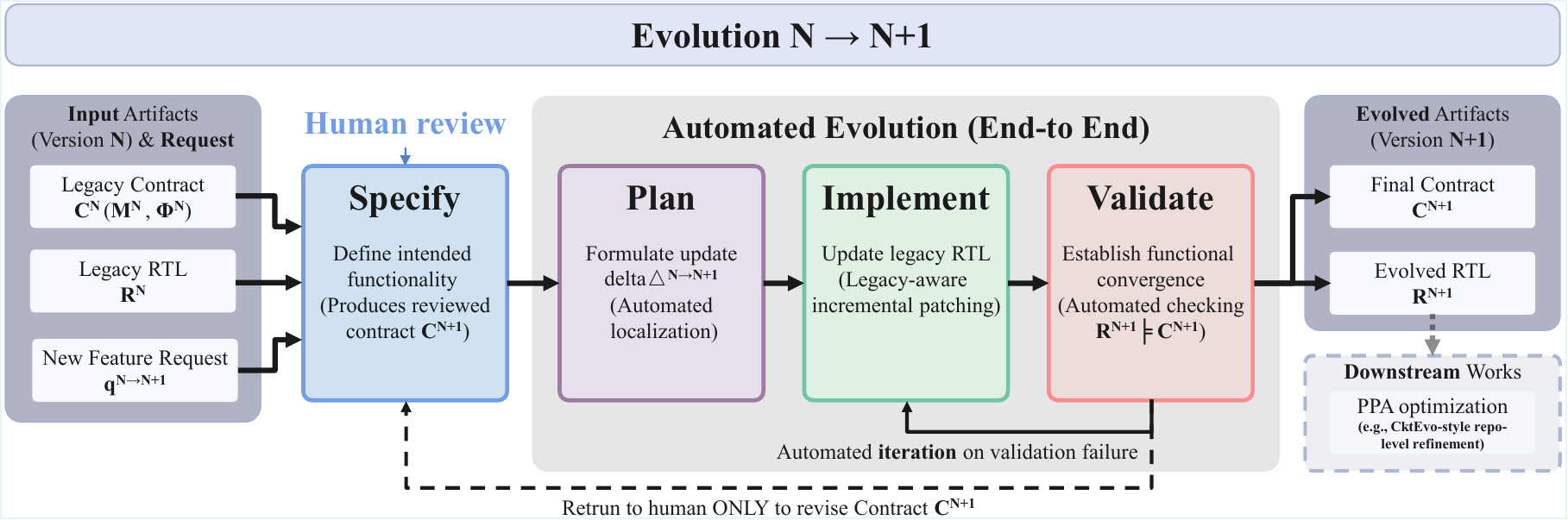}
    \vspace{-20pt}
    \caption{Spec-driven hardware evolution workflow. Human review is required only for contract construction and revision. Once the next-version contract is approved, planning, RTL update, and validation proceed automatically in a legacy-aware iterative loop.}
    \label{fig:workflow}
\end{figure}

\section{Problem Formulation and Workflow}
\label{sec:problem}

We formulate \emph{spec-driven hardware evolution} as the problem of evolving a validated legacy RTL design from version $N$ to version $N\!+\!1$ under a reviewed executable contract for the next version.

Let
\begin{equation}
\mathcal{C}^{v} = (\mathcal{M}^{v}, \Phi^{v})
\end{equation}
denote the executable contract of version $v$, where $\mathcal{M}^{v}$ is a behavior-level executable reference and $\Phi^{v}$ specifies the contract semantics, including interface conventions, observation rules, timing alignment, and checking conditions. Let $R^{v}$ denote the RTL implementation of version $v$. In our current instantiation, $\mathcal{M}^{v}$ is realized as an executable software reference model.

Given a new feature request $q^{N \rightarrow N+1}$, the goal is to construct an updated contract $\mathcal{C}^{N+1}$ together with an evolved RTL implementation $R^{N+1}$ such that the new implementation satisfies the externally visible behavior required by $\mathcal{C}^{N+1}$. The contract is not an implementation recipe; rather, it defines \emph{what} the next version must satisfy, while leaving \emph{how} the change is realized in RTL to the evolution process.

Each contract $\mathcal{C}^{v}$ induces a set of legal transactions $\mathcal{T}^{v}$. For a transaction $\tau \in \mathcal{T}^{v}$, let $Y_{\mathcal{C}}(\tau)$ and $Y_{R}(\tau)$ denote the observable outcomes produced by the contract and the RTL, respectively, under the same contract-defined interaction. We say that $R^{v}$ is \emph{transactionally equivalent} to $\mathcal{C}^{v}$ if and only if
\begin{equation}
\forall \tau \in \mathcal{T}^{v}, \quad
Y_{R}(\tau) = Y_{\mathcal{C}}(\tau).
\label{eq:transactional_equiv}
\end{equation}
Here, the observable outcome includes only contract-defined externally visible behavior and may span multiple RTL cycles according to the timing semantics in $\Phi^{v}$. In our setting, Eq.~(\ref{eq:transactional_equiv}) can be established through formal verification and, when appropriate, co-simulation between the executable reference and RTL.

Accordingly, the correctness objective of hardware evolution is
\begin{equation}
R^{N+1} \models \mathcal{C}^{N+1},
\label{eq:contract_sat}
\end{equation}
where $\models$ denotes contract satisfaction under the transactional semantics of Eq.~(\ref{eq:transactional_equiv}).

Unlike one-shot RTL generation, our setting explicitly exploits validated legacy artifacts. Let
\begin{equation}
\Delta^{N \rightarrow N+1} = \mathrm{Diff}(\mathcal{C}^{N}, \mathcal{C}^{N+1})
\end{equation}
denote the semantic delta between the legacy and next-version contracts. This delta identifies the intended behavioral change introduced by the new version and serves as the basis for subsequent planning, localized RTL update, and validation. The evolved design is therefore obtained by incrementally modifying the trusted legacy implementation $R^{N}$ under the guidance of $\Delta^{N \rightarrow N+1}$, rather than by regenerating RTL from scratch.

A key aspect of the formulation is the division between human review and automation. In \emph{Specify}, the request $q^{N \rightarrow N+1}$ is refined into a candidate contract $\mathcal{C}^{N+1}$, which must be reviewed and approved by a human designer because it defines the semantic target of version $N\!+\!1$. Once approved, the remaining stages proceed automatically. Specifically, \emph{Plan} derives the cross-version semantic delta and localizes the RTL regions most relevant to the intended change; \emph{Implement} updates the validated legacy RTL in a legacy-aware manner; and \emph{Validate} checks whether the candidate implementation satisfies Eq.~(\ref{eq:contract_sat}) through proof-oriented checking and/or co-simulation. If validation fails, the system iterates automatically over update and repair, and returns to human review only when the contract itself must be revised.

Therefore, as shown in Fig.~\ref{fig:workflow}, \emph{spec-driven hardware evolution} is a hybrid workflow with a human-reviewed contract-construction front end and an automated hardware-evolution back end. The target is not a standalone RTL artifact, but a contract-satisfying next-version implementation obtained through legacy-aware, proof-guided evolution.

%% file: tex/methodology.tex
\begin{figure*}[th]
    \centering
    \includegraphics[width=0.9\linewidth]{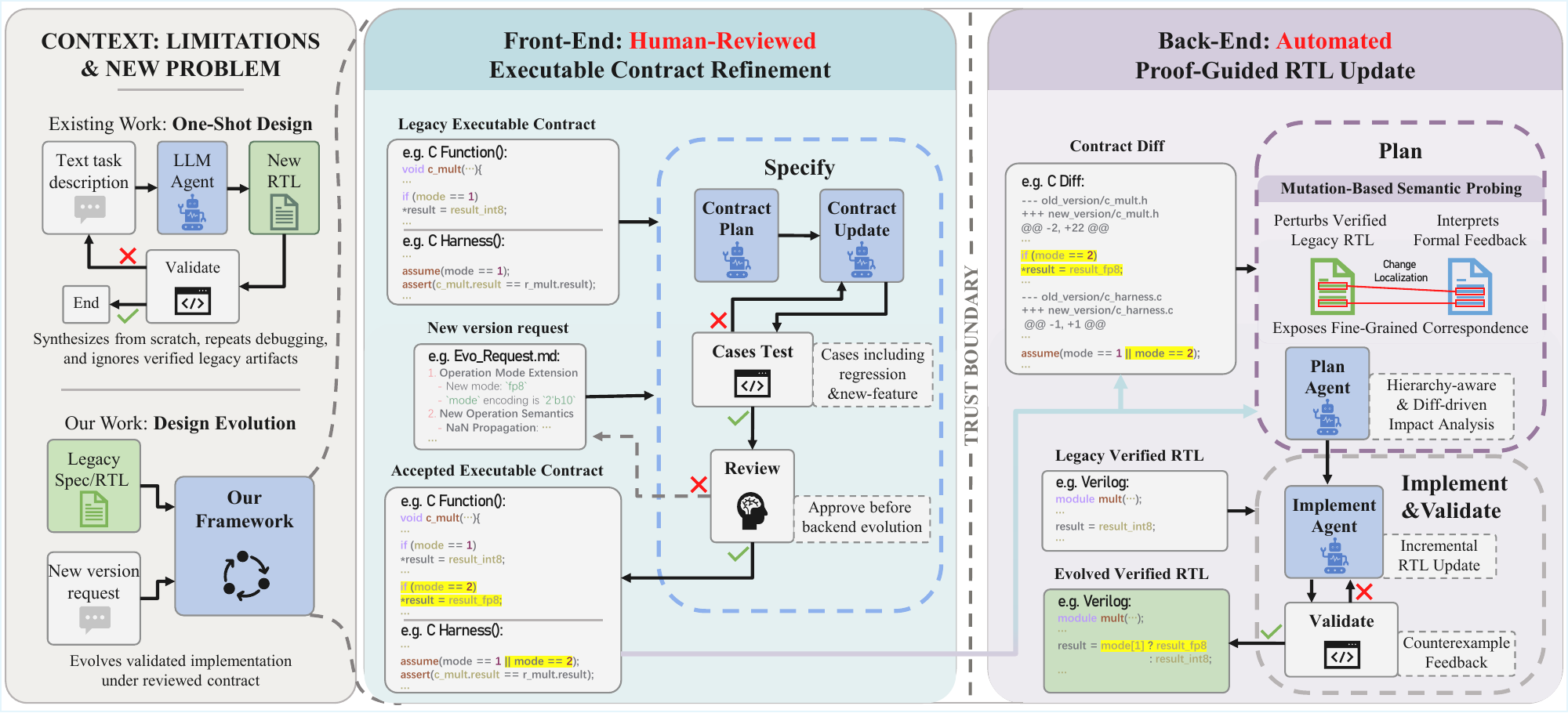}
    \vspace{-10pt}
    \caption{Overall workflow of the proposed spec-driven hardware evolution framework.}
    \label{fig:overview}
\end{figure*}

\section{Spec-Driven Hardware Evolution}
\label{sec:methodology}

\subsection{Overview}

This section describes how the proposed framework realizes \emph{spec-driven hardware evolution} in practice. Given a legacy design pair $(\mathcal{C}^{N}, R^{N})$ and a feature request $q^{N \rightarrow N+1}$, the goal is to construct an evolved implementation $R^{N+1}$ that satisfies the reviewed next-version contract $\mathcal{C}^{N+1}$. As shown in Fig.~\ref{fig:overview}, the framework follows the four stages introduced earlier: \emph{Specify}, \emph{Plan}, \emph{Implement}, and \emph{Validate}.

In \emph{Specify}, the legacy contract and the evolution request are refined into $\mathcal{C}^{N+1} = (\mathcal{M}^{N+1}, \Phi^{N+1})$, where $\mathcal{M}^{N+1}$ is written in C/C++ to define the required behavior and $\Phi^{N+1}$ captures the interface, observation, and timing semantics used in downstream checking. In \emph{Plan}, the framework analyzes the semantic delta $\Delta^{N \rightarrow N+1}$ and localizes the RTL regions most relevant to the intended change, aided by mutation-based semantic probing. In \emph{Implement}, it performs legacy-aware RTL update on top of $R^{N}$ to construct candidate versions of $R^{N+1}$. In \emph{Validate}, each candidate is checked against $\mathcal{C}^{N+1}$ using \texttt{hw-cbmc}~\cite{mukherjee2017formal}, where harness functions encode the functional and timing requirements in $\Phi^{N+1}$. Counterexamples are then used to guide repair and re-validation until contract satisfaction is established or the contract itself must be revised.

The following subsections detail these stages and the supporting mechanisms that improve the stability and efficiency of the overall workflow.

\subsection{Specify: Executable Contract Refinement}

To realize the \emph{Specify} stage, we introduce an \emph{executable contract refinement} process that evolves the contract from version-N to version-N+1. The process takes as input the legacy executable reference $\mathcal{M}^{N}$ together with the evolution request $q^{N \rightarrow N+1}$, and produces an approved next-version contract $\mathcal{C}^{N+1} = (\mathcal{M}^{N+1}, \Phi^{N+1})$. 

In our implementation, $\mathcal{M}^{N+1}$ is written in C/C++ to define the required behavior at the executable-reference level, rather than to prescribe an implementation for synthesis. The accompanying contract semantics $\Phi^{N+1}$ are realized through \emph{harness functions} used by \texttt{hw-cbmc}. Concretely, these harness functions encode three aspects of the contract: \texttt{assume} clauses constrain the admissible input space and environment conditions, \texttt{next\_timeframe()} specifies the intended temporal alignment between the executable reference and the RTL, and \texttt{assert} clauses define the required relations between contract-level and RTL-level observable outputs. Together, $\mathcal{M}^{N+1}$ and $\Phi^{N+1}$ specify \emph{what} the next version must satisfy, while leaving the RTL realization to later stages.

This stage is necessary because the evolution request alone is typically incomplete and cannot serve as a sufficiently precise semantic target for automated RTL update. A fully manual alternative, in which an engineer directly edits both the executable reference and the harness-level checking semantics, is possible but costly, and it makes completeness and consistency harder to maintain. More importantly, without an explicit and reviewable contract, downstream stages cannot cleanly distinguish ambiguity in intended next-version behavior from errors in RTL realization. We therefore place contract refinement before automated RTL evolution and keep it human-reviewed: automation improves efficiency and consistency, while human judgment remains responsible for resolving residual ambiguity and approving the final contract.

Within the front-end \emph{Specify} stage, contract refinement is carried out in an isolated workspace through four subprocesses: \emph{Contract Planning}, \emph{Contract Update}, \emph{Contract Testing}, and \emph{Human Review}. These subprocesses are specific to contract construction and should be distinguished from the RTL-side \emph{Plan}--\emph{Implement}--\emph{Validate} loop described below. \emph{Contract Planning} analyzes the legacy executable reference $\mathcal{M}^{N}$ together with the request $q^{N \rightarrow N+1}$ and produces a structured update plan covering interface changes, intended feature updates, and affected behavioral regions. \emph{Contract Update} then incrementally revises a copy of the legacy executable reference and its associated harness functions, rather than rebuilding the contract from scratch, so that unchanged behavior remains stable while the semantic delta becomes explicit in $\mathcal{C}^{N+1}$.

The candidate contract is then checked in \emph{Contract Testing}. At this stage, the updated executable reference and harness functions are exercised under broad-coverage tests, including both regression cases for preserved version-$N$ behavior and new-feature cases for the modified version-$N{+}1$ behavior. In addition, the harness is validated for contract soundness: \texttt{assume} clauses are examined to ensure that the intended input domain is neither under-constrained nor spuriously restrictive, \texttt{next\_timeframe()} calls are checked to ensure that the temporal semantics are aligned with the intended hardware behavior, and \texttt{assert} clauses are checked to ensure that the observable relation between the executable reference and RTL is stated completely and unambiguously. When tests fail, the system performs localized repair and re-validation; when ambiguity remains unresolved within the iteration budget, the issue is escalated to the engineer for clarification or contract refinement.

Finally, in \emph{Human Review}, the resulting contract $\mathcal{C}^{N+1}$ is reviewed before release with emphasis on functional completeness, contract verifiability, and backward compatibility. In particular, the review checks whether the executable reference captures the intended next-version behavior, whether the harness functions faithfully encode the corresponding input, timing, and output semantics, and whether the resulting contract is suitable for downstream proof-oriented checking with \texttt{hw-cbmc}. Only after this approval does the workflow proceed to the automated hardware-evolution back end.

\begin{figure}[t]
    \centering
    \includegraphics[width=0.9\columnwidth]{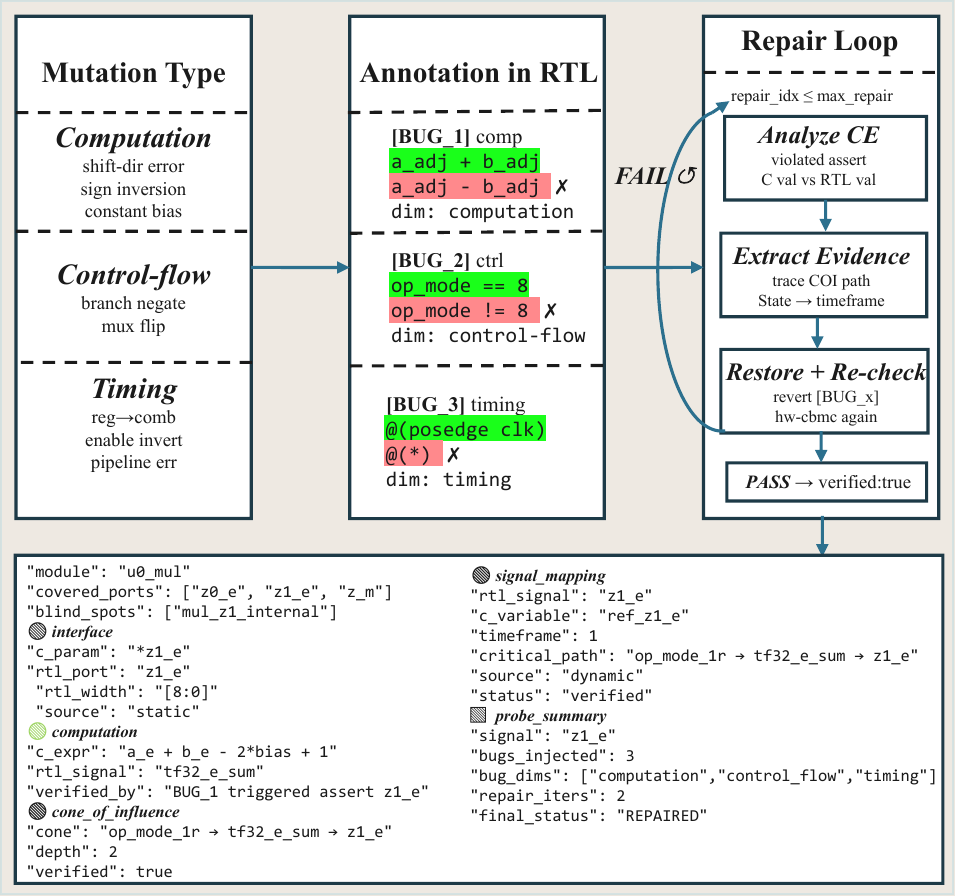}
    \vspace{-10pt}
    \caption{Mutation-based semantic probing procedure and \texttt{contract\_rtl\_map} output.}
    \label{fig:probing}
\end{figure}

\subsection{Plan: Mutation-Based Semantic Probing and Change Localization}
\label{sec:probing}

To realize the \emph{Plan} stage, we formulate planning as a problem of \emph{semantic alignment} and \emph{change localization}. Given the legacy contract $\mathcal{C}^{N} = (\mathcal{M}^{N}, \Phi^{N})$, the approved next-version contract $\mathcal{C}^{N+1}$, and the legacy RTL $R^{N}$, this stage first derives the cross-version semantic delta $\Delta^{N \rightarrow N+1} = \mathrm{Diff}(\mathcal{C}^{N}, \mathcal{C}^{N+1})$, and then produces a structured \texttt{contract\_rtl\_map} recording verified correspondences between the behavior defined by $\mathcal{M}^{N}$ and the RTL regions in $R^{N}$ that realize it. Together with $\Delta^{N \rightarrow N+1}$, this mapping provides the localization basis for the subsequent \emph{Implement} stage.

To build \texttt{contract\_rtl\_map}, we introduce \emph{mutation-based semantic probing}, which injects controlled perturbations into $R^{N}$ and interprets the resulting verification feedback under $\Phi^{N}$ as evidence for how the behavior specified by $\mathcal{M}^{N}$ is realized in the legacy RTL. Fig.~\ref{fig:probing} illustrates the procedure with a concrete example from \texttt{u0\_mul}.

Static analysis alone can only provide candidate correspondences: it cannot determine which RTL regions are exercised under $\Phi^{N}$, nor confirm that an inferred correspondence is valid under proof-oriented checking. Mutation-based semantic probing addresses this by treating the verifier as a semantic oracle and extracting alignment evidence from controlled perturbation and verification response.

\textbf{Static Analysis.}
The procedure begins with a static read of $\mathcal{M}^{N}$ and $R^{N}$ to establish an initial correspondence structure across five dimensions: interface mapping, computation mapping, control-flow mapping, timing mapping, and cone-of-influence mapping. Cone-of-influence mapping plays a central role in later fault localization, as it identifies the signals and logic regions contributing to each observed outcome in $\mathcal{M}^{N}$. All entries are initially marked \texttt{verified:false}, indicating structural hypotheses pending dynamic confirmation.

\textbf{Mutation and Probing.}
For each observed target derived from $\Phi^{N}$, we generate mutant
RTL variants by injecting controlled datapath-level faults across
multiple mapping dimensions. Each mutant perturbs computation,
control-flow, and timing aspects simultaneously to elicit richer proof
feedback than single-dimension perturbations, without over-constraining
any one dimension. Supported fault types include shift-direction errors,
arithmetic sign inversions, and conditional branch negations, among
others. Each mutant is checked by \texttt{hw-cbmc} against
$\mathcal{M}^{N}$ under $\Phi^{N}$. Signals that are successfully
exercised and validated are recorded in \texttt{covered\_ports};
those that remain insensitive to repeated perturbations are classified
as \emph{blind spots}.

\textbf{Restorative Repair.}
When verification fails on a mutated variant, the \texttt{hw-cbmc} counterexample is analyzed to validate or reject the current correspondence hypothesis. Unlike the repair in the \emph{Validate} stage, the purpose here is not to repair RTL for release, but to determine whether the hypothesized correspondence between $\mathcal{M}^{N}$ and $R^{N}$ is supported by proof feedback. This step extracts dynamic alignment evidence unavailable from static analysis, including timeframe sensitivity and critical propagation paths. Minimal restorative edits are applied along the cone-of-influence path to remove the injected perturbation and re-check the correspondence under $\Phi^{N}$. Confirmed entries are promoted to \texttt{verified:true}; insensitive signals are added to \texttt{blind\_spots}. Blind spots are propagated to the localization guidance as low-confidence regions, where the planner conservatively preserves existing logic and flags them for additional scrutiny during the \emph{Validate} stage.

Upon completion, the planner derives localization guidance from
\texttt{contract\_rtl\_map} and $\Delta^{N \rightarrow N+1}$,
identifying where legacy logic should be preserved and where
modification effort should be concentrated. The resulting
\texttt{contract\_rtl\_map} also records verified correspondences
between $\mathcal{M}^{N}$ and $R^{N}$, together with critical paths,
\texttt{covered\_ports}, and blind-spot annotations, serving as the
semantic grounding for the subsequent \emph{Implement} stage.

\begin{algorithm}[t]
\footnotesize
\caption{Mutation-Based Semantic Probing}
\label{alg:probing}
\begin{algorithmic}[1]
\REQUIRE legacy contract $C^N = (\mathcal{M}^N, \Phi^N)$, approved contract $C^{N+1}$,
    \\\hspace*{2.4em} legacy RTL $R^N$
\ENSURE \texttt{contract\_rtl\_map}, \texttt{blind\_spots},
        localization guidance
\STATE $\Delta^{N \rightarrow N+1} \leftarrow
       \mathrm{Diff}(\mathcal{C}^{N}, \mathcal{C}^{N+1})$
\STATE $\texttt{contract\_rtl\_map} \leftarrow
       \textsc{StaticAnalysis}(\mathcal{M}^{N}, R^{N})$
\FOR{each observed target $s$ derived from $\Phi^{N}$}
    \STATE $\mu \leftarrow \textsc{InjectFaults}(R^{N}, s)$
    \IF{$\texttt{hw-cbmc}(\mu,\ \mathcal{M}^{N},\ \Phi^{N}) = \textsc{Fail}$}
        \STATE $e \leftarrow \textsc{AnalyzeCounterexample}(\mu)$
        \STATE $\texttt{contract\_rtl\_map} \leftarrow
               \textsc{RestorativeRepair}(\mu,\ e,
               \texttt{contract\_rtl\_map})$
    \ELSE
        \STATE $\texttt{blind\_spots} \leftarrow
               \texttt{blind\_spots} \cup \{s\}$
    \ENDIF
\ENDFOR
\STATE $\text{guidance} \leftarrow
       \textsc{Derive}(\texttt{contract\_rtl\_map},\
       \Delta^{N \rightarrow N+1})$
\RETURN $\texttt{contract\_rtl\_map},\ \texttt{blind\_spots},\
         \text{guidance}$
\end{algorithmic}
\end{algorithm}

\subsection{Implement and Validate: Proof-Guided RTL Update Loop}

Given the reviewed contract $\mathcal{C}^{N+1}$, the validated legacy RTL $R^{N}$, the cross-version semantic delta $\Delta^{N \rightarrow N+1}$, and the \texttt{contract\_rtl\_map} produced in the \emph{Plan} stage, hardware evolution proceeds to the \emph{Implement} and \emph{Validate} stage. The objective is not to obtain the final RTL in a single pass, but to iteratively construct and refine a candidate $R^{N+1}$ until it satisfies $\mathcal{C}^{N+1}$. In this sense, the core of the automated back end is \emph{proof-guided closure}: candidate RTL is updated under localized guidance, checked against the reviewed contract, and repaired using verification feedback until convergence, budget exhaustion, or escalation to contract revision.

\textbf{Implement.}
The system applies the structured modification guidance from \emph{Plan} to a working copy of $R^{N}$ in order to construct a candidate $R^{N+1}$. Rather than regenerating hardware description from scratch, this stage performs localized, legacy-aware patching on top of the trusted implementation substrate. Previously validated logic is preserved whenever possible, and edits are constrained to the modules, signals, and code regions implicated by $\Delta^{N \rightarrow N+1}$ and \texttt{contract\_rtl\_map}. As a result, implementation remains localized in both functional scope and structural impact.

\textbf{Validate.}
The candidate $R^{N+1}$ is checked against $\mathcal{C}^{N+1}$ using \texttt{hw-cbmc}. In our implementation, the executable reference $\mathcal{M}^{N+1}$ and the candidate $R^{N+1}$ are linked through harness functions that encode the contract-defined functional and timing requirements in $\Phi^{N+1}$. Verification success is treated as functional convergence under the next-version contract. If contract satisfaction is not established, the verifier returns structured outcomes, including counterexamples and related diagnostic evidence, which are passed to the repair step to localize the source of divergence and guide targeted correction.

\textbf{Repair and Closure.}
When verification fails, the system reuses the counterexample-analysis backend together with the semantic correspondences established during \emph{Plan}, but now with a different objective: not to refine alignment between $\mathcal{M}^{N}$ and $R^{N}$, but to repair the candidate $R^{N+1}$ itself. Counterexamples are used to localize faulty modules, signals, and control/data paths in the candidate $R^{N+1}$, after which the system applies targeted fixes and re-enters validation. This implement--validate--repair loop continues until one of three stopping conditions is reached: verification succeeds; the observed failures indicate that the contract itself requires clarification or revision; or a predefined iteration or cost budget is exhausted.

In this way, LLM-driven RTL modification is embedded within a formally checkable and iteratively convergent refinement loop. The resulting workflow prioritizes contract satisfaction through localized legacy-aware evolution, rather than unconstrained one-shot generation.

\subsection{Supporting Mechanisms}

Beyond the stage-wise workflow above, the framework relies on several cross-cutting mechanisms that improve stability, repair quality, and execution efficiency.

\textbf{Task-Isolated Subagents.}
Hardware evolution is a long-horizon and tightly constrained generation-and-verification task. As context accumulates, a single agent is prone to instruction drift, which gradually blurs stage boundaries and degrades both stability and correctness. To address this issue, we adopt a task-isolated multi-agent design in which planning, implementation, validation, and repair are executed by fresh subagents, while a main agent is responsible only for orchestration and artifact passing. This design shortens the effective context span, reduces instruction drift, and stabilizes long-horizon hardware evolution under strong procedural constraints.

\textbf{Hierarchy-Aware Bottom-Up Repair.}
In hierarchical hardware designs, local defects in lower-level modules often propagate upward and eventually manifest as failures at top-level verification. Directly repairing the top-level module in such cases can misidentify the root cause and introduce ineffective changes. We therefore explicitly extract the design hierarchy and perform hierarchy-aware bottom-up repair: a failed module is repaired only after its dependent submodules have been stabilized. This strategy respects the actual dependency structure of the design, aligns repair order with error-propagation paths, and reduces mislocalized fixes caused by upstream failure propagation.

\textbf{Scripted Verification Orchestration.}
The validation stage involves repetitive operations, including constructing verification commands, invoking \texttt{hw-cbmc}, collecting results, and generating structured logs. Fully relying on agent reasoning for these steps is inefficient and incurs unnecessary token overhead. We therefore encapsulate verification-environment parsing, command construction, batch execution, and report generation into Python scripts, which the agent invokes as standardized verification backends. This scripted design improves execution efficiency, consistency, and reproducibility, while reducing avoidable reasoning cost during validation.

%% file: tex/experiment.tex
\section{Experiments}
\subsection{Experimental Setup}
Existing public benchmarks such as VerilogEval~\cite{liu2023verilogeval}, RTLLM~\cite{lu2024rtllm}, and CktEvo~\cite{shi2026cktevo} do not match our setting, because they target either prompt-to-RTL generation or function-preserving optimization, rather than \textbf{spec-driven functional evolution} from trusted legacy RTL under aligned executable contracts. A realistic evaluation in our setting requires three key assets: a verified legacy RTL, a semantically aligned next-version executable contract, and proof-consumable change requests. We therefore construct controlled version-evolution tasks based on \textbf{dot\_core}, a TPU datapath block~\cite{jouppi2017datacenter} originally supporting INT8, FP4, and FP8 operations, which implements a 32-way dot-product accumulation pipeline across a multi-level module hierarchy.

The benchmark follows a \textbf{version-N~$\to$~version-N+1} trajectory, where version-N+1 adds \textbf{TF32 support}~\cite{valero2023mixed} by modifying
mantissa and exponent widths across the multiplier and alignment datapaths and propagating interface changes through the hierarchy. Table~\ref{tab:benchmark} summarizes the resulting benchmark statistics at both the top-module and representative submodule levels, with approximately \textbf{two person-weeks} of engineering effort required for benchmark construction.

On the contract side, the $\mathcal{C}^{N+1}$ entries reported in Table~\ref{tab:benchmark} are produced by the front-end \emph{Specify} process, which incrementally updates the executable reference and associated checking semantics to reflect the next-version behavior. In our benchmark construction, this front-end update remains lightweight: once the intended TF32 behavior is specified, preparing and reviewing the corresponding contract typically requires only about \textbf{one person-day} of engineering effort. This is substantially smaller than the cost of backend RTL evolution and formal closure, and is consistent with our intended usage scenario in which the contract serves as a reviewed semantic target before automated hardware evolution begins.

All experiments in this section are therefore conducted under a fixed human-reviewed next-version executable contract. We treat contract construction as a separate front-end process and exclude the \textit{Specify} stage from the main quantitative evaluation, so that the reported results focus on the end-to-end automated evolution capability of the \emph{Plan}--\emph{Implement}--\emph{Validate} backend.

\begin{table}[t]
\centering
\caption{Benchmark statistics for dot\_core version evolution tasks.}
\vspace{-10pt}
\label{tab:benchmark}
\resizebox{\columnwidth}{!}{
\begin{tabular}{lcccc}
\toprule
Module & $R^{N}$ & $R^{N+1}$ ($\Delta$LoC) & $\mathcal{C}^{N}$ & $\mathcal{C}^{N+1}$ ($\Delta$LoC) \\
\midrule
dot\_core & 678 & 678 (+65,-65) & 232 & 233 (+34,-33) \\
u0\_mul & 331 & 336 (+87,-82) & 215 & 319 (+109,-5) \\
u1\_exp\_align & 215 & 219 (+59,-55) & 111 & 135 (+40,-16) \\
u2\_add\_vector & 333 & 366 (+105,-72) & 288 & 330 (+63,-21) \\
u3\_add & 196 & 196 (--) & 183 & 183 (--) \\
u4\_add\_mix & 214 & 214 (+4,-4) & 316 & 316 (+2,-2) \\
u5\_norm & 111 & 111 (--) & 172 & 172 (--) \\
u6\_ffp2fp\ & 109 & 109 (+5,-5) & 106 & 106 (+2,-2) \\
\bottomrule
\end{tabular}
}
\vspace{-10pt}
\end{table}


Our evaluation uses a workstation equipped with an AMD Ryzen Threadripper PRO 9985WX 64-core CPU. We use \texttt{hw-cbmc}~\cite{mukherjee2017formal} as the primary verification backend for bit-exact equivalence checking between $R^{N+1}$ and $\mathcal{C}^{N+1}$. For QoR evaluation, we synthesize the evolved RTL using \texttt{Synopsys Design Compiler}\footnote{\url{https://www.synopsys.com/implementation-and-signoff/rtl-synthesis-test/design-compiler.html}} with the \texttt{ASAP7 7.5-track standard cell library}~\cite{8203889}. Since the goal of this work is functional convergence rather than physical design optimization, post-synthesis cell area and critical-path delay are reported only as secondary QoR metrics.

\subsection{Main Results Across Models}

We evaluate the proposed workflow on the full \texttt{dot\_core}
evolution task across different LLMs, using the same
OpenCode\footnote{\url{https://opencode.ai/}}-based execution interface throughout. As noted above, the next-version executable contract is fixed and human-reviewed before backend automation begins. In particular, the contract-side updates summarized in Table~\ref{tab:benchmark} are obtained by the front-end \emph{Specify} process and require only modest engineering review effort. The experiments in this subsection therefore measure the end-to-end automated performance of the backend workflow, rather than the cost of contract construction itself.

All comparisons in this section use the same workflow configuration, skill implementation, verification backend, and budget setting. Here, a \emph{run} denotes one complete backend evolution attempt, from candidate RTL generation to either successful top-level \texttt{dot\_core} verification or budget exhaustion.

\begin{figure}[t]
    \centering
    \includegraphics[width=0.95\linewidth]{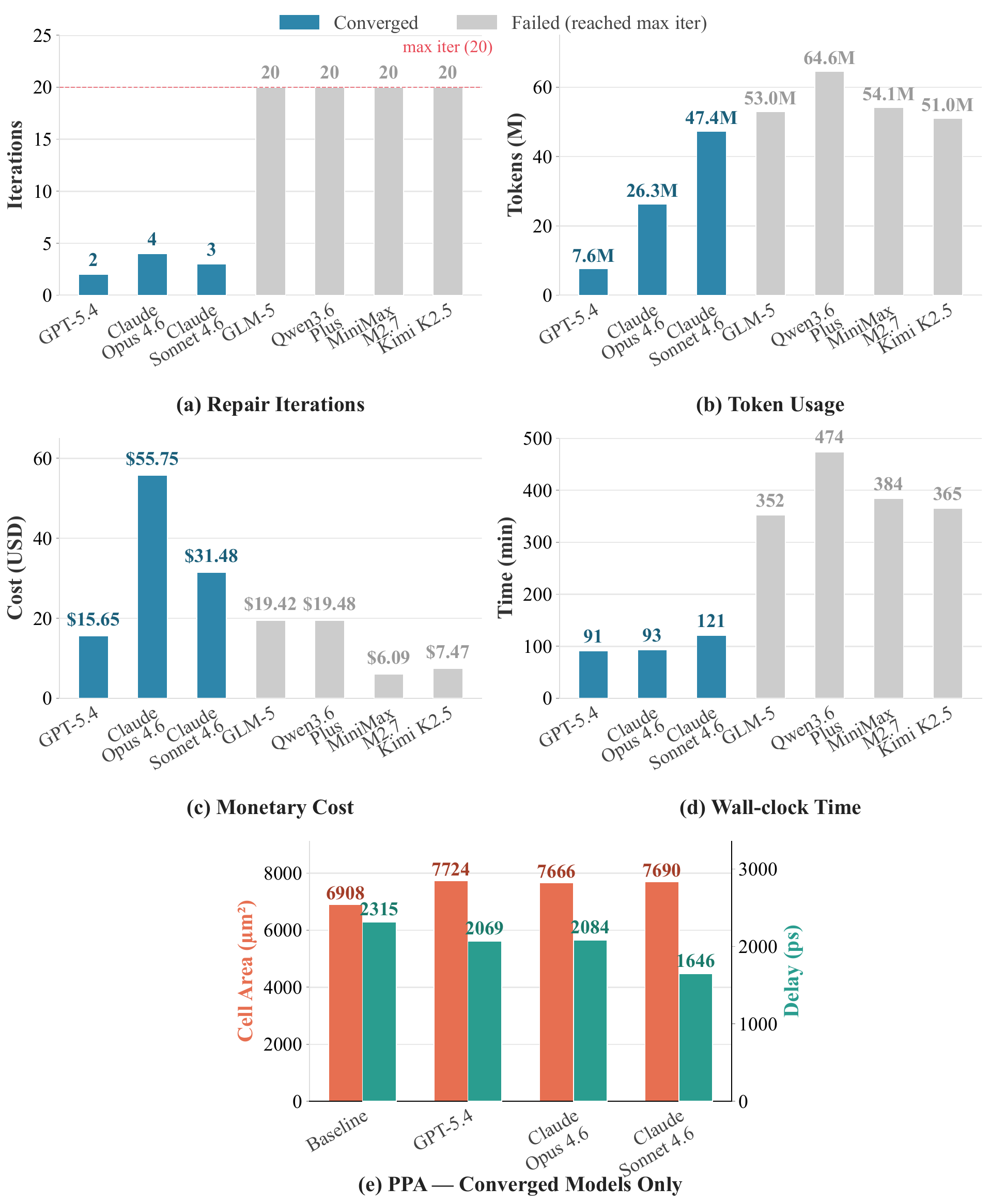}
    \vspace{-10pt}
\caption{Comparison of repair iterations, token usage, monetary cost, wall-clock time, and post-synthesis PPA across models on the \texttt{dot\_core} evolution task.}
    \vspace{-10pt}
    \label{fig:model_comparison}
\end{figure}

Fig.~\ref{fig:model_comparison} reports the results across models,
including convergence, repair iterations, token usage, monetary cost,
wall-clock time, and post-convergence PPA metrics. Since the goal of
this work is functional convergence rather than explicit PPA
optimization, Area and Timing are reported as secondary metrics
relative to the legacy RTL baseline.

Based on these results, we select GPT-5.4 as the default model for the subsequent architecture-comparison and ablation experiments. The rationale is threefold:

\textbf{Absolute Task Capability:} The hardware evolution task presents significant complexity. Within the maximum allowed iterations (20), only top-tier models (GPT-5.4, Claude Opus 4.6, and Claude Sonnet 4.6) successfully converged and passed the validation. The other tested models failed to realize the correct RTL updates, narrowing our baseline candidates strictly to the successful ones.

\textbf{Superior Evolution Efficiency:} Among the successful candidates, GPT-5.4 demonstrates the highest efficiency in driving functional convergence. It achieved a passing result in only 2 iterations and required the shortest execution time (91 minutes), outperforming both Claude Opus 4.6 (4 iterations, 93 minutes) and Claude Sonnet 4.6 (3 iterations, 121 minutes). This indicates that GPT-5.4 possesses stronger reasoning capabilities for localizing and updating  $R^{N}$, requiring fewer corrective feedback loops.

\textbf{Optimal Resource and Cost Effectiveness:} Given that subsequent ablation studies require extensive repetitive experiments, token consumption and financial cost are critical factors. GPT-5.4 is exceptionally token-efficient, consuming only 7.6M tokens—roughly 29\% of Opus 4.6 and 16\% of Sonnet 4.6. Consequently, it offers the lowest cost (\$15.65) among the successful models.

In summary, GPT-5.4’s comprehensive advantages in functional correctness, iteration speed, and operational cost make it the most robust and practical foundation for our framework evaluation

\subsection{Fixed vs. Flexible Workflow Orchestration}
Using the default model selected above, we compare two workflow organizations on the full end-to-end \texttt{dot\_core} evolution task. Both variants implement the same core task stages, but differ in how these stages are organized and coordinated:

\begin{enumerate}
    \item \emph{LangGraph-based}: hardens the stages into a fixed multi-agent execution graph with predefined inter-agent communication and execution order.
    \item \emph{Engineer-skill workflow} (ours): uses a skill-driven multi-agent framework with explicit stage decomposition but more flexible coordination.
\end{enumerate}

\begin{table}[t]
\centering
\caption{Comparison between fixed-pipeline and skill-driven orchestration on \texttt{dot\_core} evolution.}
\label{tab:arch}
\resizebox{\columnwidth}{!}{
\begin{tabular}{llccccccc}
\toprule
Workflow & Result & Iter. & Tokens & Cost & Time(min) & Cell Area($\mu$m$^2$) & Delay(ps) \\
\midrule
LangGraph-based            & pass  & 5 & 914,400    & \$3.47  & 65  & 7650.96 & 2155.43 \\
Engineer-skill workflow (ours) & pass  & 2 & 7,619,937  & \$15.65 & 91  & 7723.55 & 2068.80 \\
\bottomrule
\end{tabular}
}

\end{table}

Table~\ref{tab:arch} reports convergence, repair iterations, and resource cost for the two workflow organizations.
The two variants exhibit a clear trade-off between workflow specification and execution flexibility. The LangGraph-based pipeline achieves lower token cost and shorter runtime in the reported run, indicating that a fixed execution structure can be highly efficient when the intermediate trajectory aligns with the predefined workflow. In contrast, the engineer-skill workflow incurs higher token consumption and wall-clock time, but converges in fewer repair iterations.

Rather than suggesting that one framework uniformly dominates the other, this comparison highlights two distinct orchestration regimes. The LangGraph-based variant hardens parts of the workflow into predefined nodes and fixed inter-node transitions, reducing per-run overhead when execution follows the expected path. By contrast, the engineer-skill workflow preserves more adaptive interaction with tools, intermediate artifacts, and repair feedback, at the cost of higher runtime and token usage.

\subsection{Ablation Studies}
\label{sec:ablation}

We assess the contribution of core components within the proposed workflow using the same default model-interface combination on the full end-to-end \texttt{dot\_core} task.
\begin{enumerate}[label=(\arabic*), leftmargin=*, itemsep=0pt, topsep=2pt]
    \item \emph{w/o task-isolated subagents}, which removes fresh sub-agent isolation while keeping the same overall workflow.
    \item \emph{w/o mutation-based semantic probing}, which disables mutation-guided semantic localization in the planning stage.
    \item \emph{w/o hierarchy-aware bottom-up repair}, which repairs all failing modules in a flat order without considering the design hierarchy.
    \item \emph{Direct diff w/o structured plan}, which replaces the structured modification plan with a direct diff-style text.
\end{enumerate}

\begin{figure}[t]
    \centering
    \includegraphics[width=\linewidth]{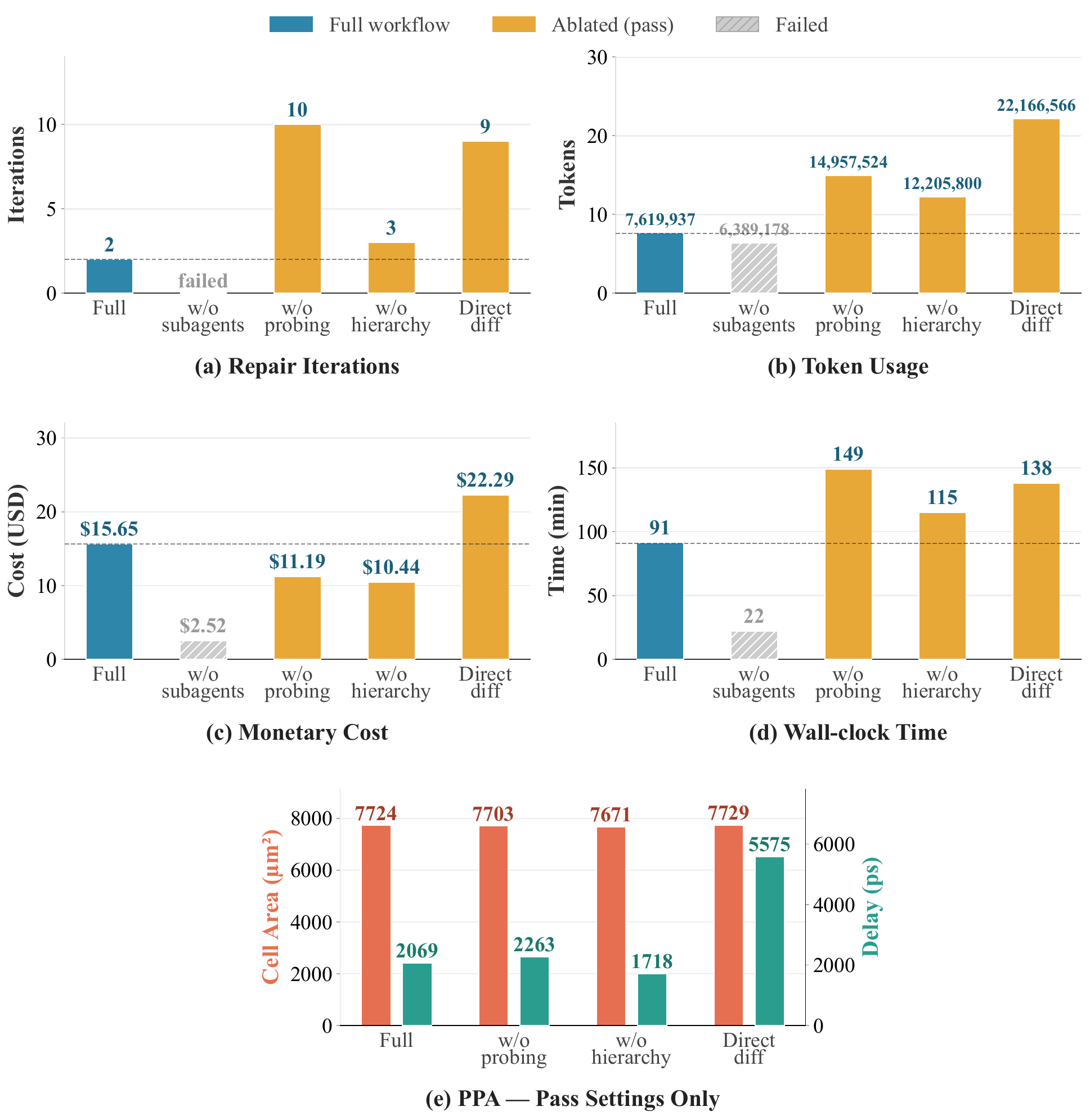}
    \vspace{-20pt}
    \caption{Ablation results on \texttt{dot\_core} evolution.}
    \label{fig:ablation}
\end{figure}

As shown in Fig.~\ref{fig:ablation}, the full workflow achieves the best overall performance, converging in only 2 iterations with lower runtime than all successful ablated variants. Variant~(1) fails outright after only two iterations, well before the maximum repair budget of 20, because the agent deviates from the prescribed workflow without fresh sub-agent isolation, confirming that task decomposition is essential for preventing instruction drift. Variant~(2) still converges but requires 10 iterations, as the \texttt{contract\_rtl\_map} that enables localized, legacy-aware patching is no longer available. Variant~(3) converges in 3 iterations but with higher token cost and runtime, since the system may misidentify root causes when submodule defects propagate to top-level verification. Variant~(4) yields the worst result among successful variants in both iteration count and token cost, indicating that raw diff information alone lacks sufficient structure to guide RTL evolution. 

In summary, the ablation results confirm that task-isolated subagents, mutation-based probing, hierarchy-aware bottom-up repair, and structured planning are complementary components whose synergy enables reliable and efficient hardware evolution.

%% file: tex/conclusion.tex
\section{Conclusion}
\label{sec:conclusion}

This paper presents \emph{spec-driven hardware evolution}, a contract-centered formulation of RTL version iteration inspired by spec-driven development. Instead of treating hardware evolution as direct prompt-to-RTL generation, we cast it as the process of evolving a validated legacy design under a reviewed executable contract for the next version. Under this formulation, the central question shifts from \emph{``what Verilog should be written?''} to \emph{``what specification should be provided, reviewed, and checked?''}. Once the intended next-version behavior is captured in an executable, verification-consumable contract, validated legacy RTL can be evolved through a structured, proof-guided process rather than regenerated from scratch.

Our results suggest that this contract-centered formulation is particularly well suited to the early stage of real hardware iteration, where the immediate objective is to reach functional convergence on revised requirements. Although our present scope focuses on functional evolution rather than PPA optimization, the evolved RTL produced by the proposed flow can naturally serve as a starting point for downstream PPA-oriented refinement. We therefore view this work not as an endpoint, but as a step toward a broader automated front-end design flow in which executable intent capture, functional convergence, and subsequent PPA optimization are integrated into a unified agentic pipeline for digital hardware development.